\documentclass[twocolumn,showpacs,preprintnumbers,amsmath,amssymb]{revtex4}

\usepackage{graphicx}
\usepackage{dcolumn}
\usepackage{bm}

\begin{document}


\title{Determination of anisotropic $H_{c2}$ up to 60 T in (Ba$_{0.55}$K$_{0.45}$)Fe$_2$As$_2$ single crystals }

\author{M. M. Altarawneh, K. Collar, and C. H. Mielke}
\affiliation{Los Alamos National Laboratory, MPA-NHMFL, Los Alamos, NM 87545}
\author{N. Ni, S. L. Bud'ko and P.C. Canfield}
\affiliation{Ames Laboratory US DOE and Department of Physics and Astronomy, Iowa State University, Ames, IA 50011}

\date{\today}

\begin{abstract}
The radio frequency penetration depth was measured in the superconductor
(Ba$_{0.55}$K$_{0.45}$)Fe$_{2}$As$_{2}$ under pulsed magnetic fields extending to 60 tesla and down to 14 K. Using these data we are able to infer a $H_{c2}(T)$, $H-T$ phase diagram, for applied fields parallel and perpendicular to the crystallographic $c$-axis. The upper critical field curvature is different for the respective orientations but they each remain positive down to 14 K. The upper critical field anisotropy is moderate, $\approx 3.5$ close to $T_c$, and drops with the decrease of temperature, reaching $\approx  1.2$ at $14 K$. These data and analysis indicate that (i) (Ba$_{0.55}$K$_{0.45}$)Fe$_{2}$As$_{2}$ anisotropy diminishes with temperature and has an unusual temperature dependence, (ii) $H_{c2} (T=0)$ for this compound may easily approach fields of 75 tesla.
\end{abstract}

\maketitle


The discovery of relatively high temperature superconductivity in several families of FeAs based compounds\cite{a,b} has raised hope in both basic and applied physics circles.  Whereas these materials present intriguing questions about mechanism, lying as they do near a confluence of structural and magnetic phase transitions,\cite{a,b,c,d} they also hold out the promise of not only high $T_c$ values but also apparently extremely high $H_{c2}$ values as well.  $RFeAs(O_{1-x}F_x)$ compounds have been shown to have $H_{c2}(T)$ curves that can be reasonably extrapolated to low temperatures $H_{c2}(0)$ values in excess of 100 T. \cite{sef,ssef} Although in some cases anisotropic $H_{c2}$ data can be inferred from magnetization measurements on polycrystalline samples \cite{poly}, to date, the lack of sizable single crystalline RFeAsO  samples has prevented thorough determinations of anisotropic $H_{c2}(T)$ curves over extended field ranges.

Fortunately, methods for growing single crystals of superconducting $(Ba_{1-x}K_{x})Fe_2 As_2$, \cite{ni} and subsequently the Sr and Ca analogues, were readily developed based on the fact that the $AFe_2 As_2$ compounds are true intermetallics and amenable to conventional metallic solution growth techniques.  The availability of large, single crystals allows for the determination of anisotropic $H_{c2}(T)$ data.  Initial, anisotropic magnetoresistivity data collected on $(Ba_{0.55}K_{0.45})Fe_2 As_2$ show that there is a very minor suppression of $T_c$ even for 14 T. \cite{ni} Extrapolations of $H_{c2}(T)$ plots inferred from these data to T = 0 K suggest that $H_{c2}(T)$  for this newly discovered superconductor may easily exceed 70 T, and interpolation of the data indicate that the upper critical field anisotropy parameter, $\gamma$ = $H_{c2}^{\perp c} / H_{c2}^{\| c}$, is approximately 3.5 close to $T_c$.  Higher field data are clearly required in both cases to improve our understanding and to place the extrapolation of $H_{c2}(T)$ on firmer footing.

	In this report we present anisotropic, high frequency susceptibility measurements on single crystals of $(Ba_{0.55}K_{0.45})Fe_2 As_2$ grown in the same manner as those used to collect the initial $H \leq 14$ T data.\cite{ni}   We are able to extend $H_{c2}(T)$  curves to significantly higher magnetic fields and lower temperatures.  As a result we find that (i) $\gamma$ varies in a continuous manner from a relative maximum of $\approx$ 3.5 near $T_c$ to a value of 1.16 near 14K, and (ii)  the extrapolation of the $H_{c2}^{\| c} (T)$ and $H_{c2}^{\perp c} (T)$ curves to lower temperature indicate that values of between  $\approx$ 75 T and 80 T are likely in this material.

	Single crystals of $(Ba_{0.55}K_{0.45})Fe_2 As_2$  were grown in precisely the manner described in reference \cite{ni}.  For these crystals the potassium content was directly measured by wavelength dispersive x-ray spectroscopy and was found to have an average value of 0.45 with a 0.07 standard deviation from layer to layer.\cite{ni}

Radio frequency (rf), contactless penetration depth measurements were performed on the single crystal sample in a 60 T pulsed field magnet with a 10 millisecond rise time and a 40 millisecond extended decay. The rf technique was used as it has proven to be a sensitive (typically $\sim$ 1 part per 1000 resolution) and accurate method for determining the upper critical field in anisotropic superconductors.\cite{chm01} A radio frequency probe based on a commercially available integrated circuit (IC) proximity detector was used to establish a high stability tank circuit oscillator. The IC delivers a voltage proportional to the tank circuit voltage. The fundamental resonant frequency is approximately 28 MHz at $T_c$. The technique is highly sensitive to small changes ($\sim 1 - 5$ nm ) in the rf penetration depth \cite{chm01} when the sample is in the superconducting state. As the magnetic field increases, the probe detects the transition to the normal state by tracking the shift in resonant frequency (proportional to the penetration depth i.e. $\Delta\lambda = \Delta f R^2 / f_0 r_s$  where $R$ is the detection coil radii, $r_s$ is the sample radii, $f_0$ is the fundamental frequency, and $\Delta f$ is the frequency shift \cite{chm01}). The IC voltage is mixed with an intermediate frequency (IF) and the difference frequency is extracted by using a low pass filter at 1.9 MHz. This simple rf technique allows for the frequency shift to be recorded with a digitizing oscilloscope at a rate of 80 ns/pt (12.5 MHz) for a duration of 100 milliseconds (1.25 M samples/channel). A simple peak finding algorithm is run on the raw data to calculate the frequency shift as a function of time. The time dependent frequency is then correlated with the applied magnetic field yielding a frequency vs. magnetic field plot (see figures 1 and 2).

\begin{figure}
\begin{center}
\includegraphics[angle=0,width=80mm]{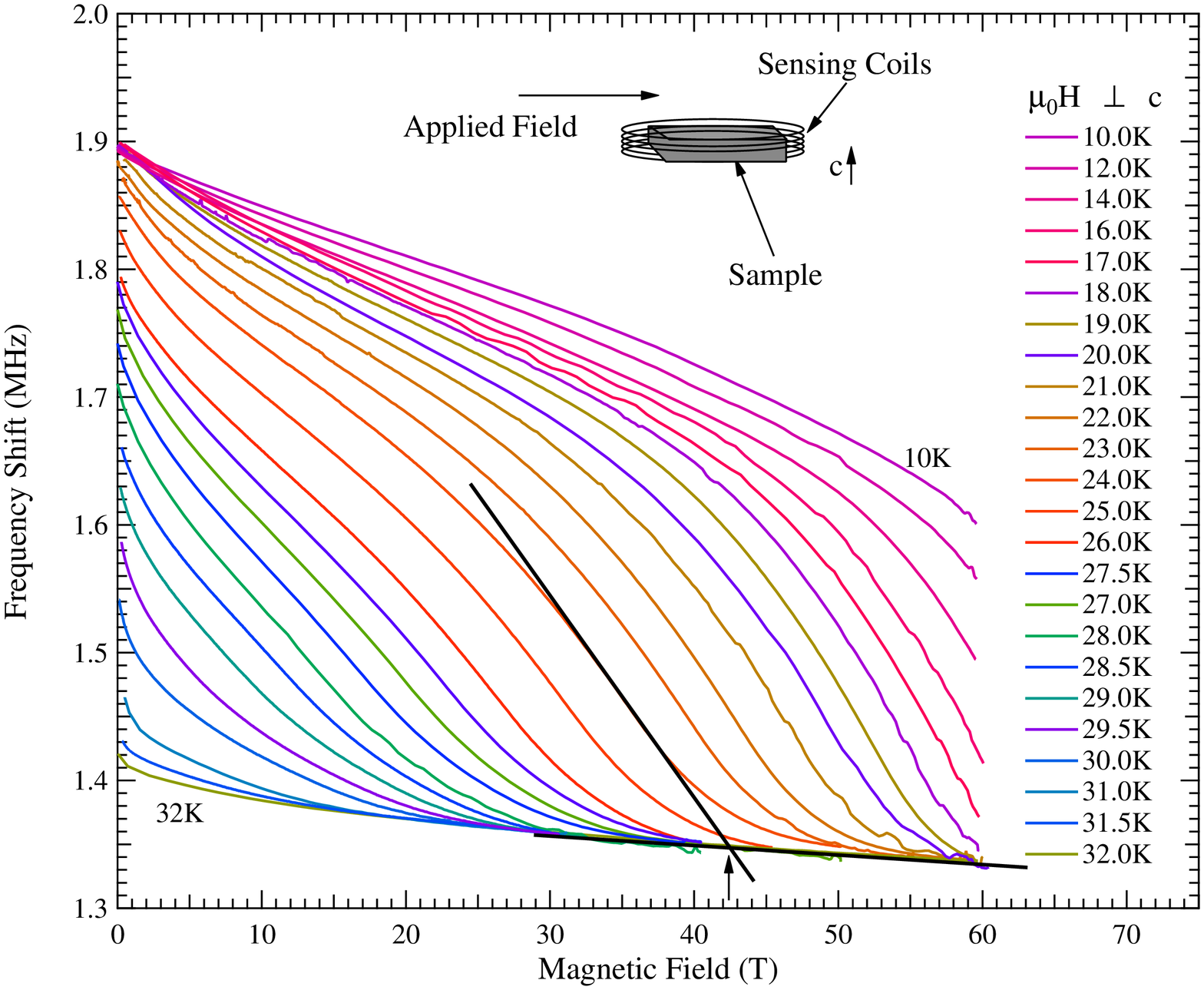}
\end{center}
\caption{(Color online) Frequency shift as a function of magnetic field applied perpendicular to the crystallographic $c$-axis at different temperatures from 10 K to 32 K. The arrow indicates the approximate $H_{c2}$ point for the 24 K field pulse and the linear guides used for determination of the point. Inset: a schematic diagram of the rf sensing coil and sample for the $\mu_0 H \perp c$ configuration. } \label{F1}
\end{figure}

A small ($0.55 \times 0.55 \times 0.014$ mm$^3$) single crystal was chosen as larger ($>1$ mm) cross section samples tend to exhibit signs of heating due to eddy currents induced in the sample, caused by the pulsed field. Our results show no hysteresis (hence no measurable heating) during the pulse indicating a good thermal anchor was established with the bath.  To determine the upper critical field anisotropy the  single crystal was measured in two orthogonal sample orientations; the applied field parallel to the conducting planes (perpendicular to the crystallographic c-axis, see inset to Fig.1) and with the applied field  normal to the conducting planes (parallel to the crystallographic c-axis, see inset to Fig.2). Alignment accuracy was within approximately 2 degrees. For applied field perpendicular to the $c$-axis (Fig.1) the sample was placed in a circular detection coil ($\approx 0.8$ mm in diameter). The sample and probe were cooled down to liquid helium temperatures inside of a double walled, non-metallic, cryostat with $\approx$ 500 torr of helium exchange gas. The onset of the superconducting transition is  clearly observed (as a resonant frequency shift) near 32 K in zero applied magnetic field (shown in inset to Fig.3).  In the second orientation ($H$ parallel to the $c$-axis) (Fig.2) a flux compensated ``figure eight'' detection coil set ($0.7$ mm diameter each) was placed with its symmetry axis parallel to the applied magnetic field. The compensated nature of the detection coil was necessary to minimize induced voltages feeding back into the rf IC. The sample was placed on the top surface of one side of the counter wound coil pair. This configuration has weaker coupling to the sample ($< 50\%$) resulting in a smaller, but still easily resolvable frequency shift. The temperature was stabilized with a conventional PID temperature controller with an accuracy of $\approx \pm 0.1$ K).

\begin{figure}
\begin{center}
\includegraphics[angle=0,width=80mm]{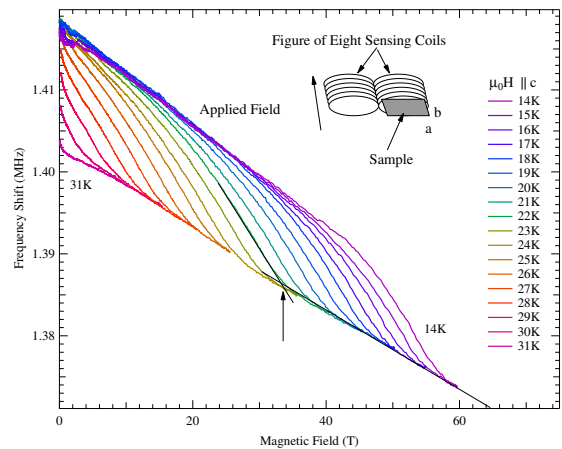}
\end{center}
\caption{(Color online) Frequency shift as a function of magnetic field applied along the crystallographic $c$-axis at different temperatures from 14 K to 31 K. Inset: a schematic diagram of the rf sensing coil and sample for the $\mu_0 H \parallel c$ configuration. } \label{F2}
\end{figure}

The criterion for determining the point at which the $H_{c2}$ transition occurs requires consistency, an no one method is entirely unambiguous.  However, the rf penetration depth technique has demonstrated a high degree of agreement with accepted thermodynamic methods,\cite{chm01}, and as a ``contactless'' method, avoids problems with contact lead problems that can be detrimental to transport measurements in pulsed magnetic fields. In addition the rf technique has the advantage of being more sensitive with a several-order-of-magnitude faster time response, hence making it suitable for millisecond duration pulsed magnetic fields (when compared to specific heat for example). Our method, for determining a consistent $H_{c2}$ point in the $\Delta f$ vs. $\mu_0 H$  data shown on figures 1 and 2, is based on identifying the point at which the slope of the rf signal (in the transition region) intercepts the slope of the normal state background. The justification is simply  that the high field  data (above $H_{c2}$)show a smooth, close to linear, magnetic field dependence. In this region the rf probe is sensitive to the normal state magnetoresistance of the sample and detection coil.  Below $H_{c2}$ the suppression of the superconducting state with increasing field leads to a field dependent frequency shift that results in a clear slope change in the data (see figures 1 and 2). The $H_{c2}$ value is determined for each subsequent fixed temperature magnetic field pulse producing the $H-T$ plot (see Fig.3).

\begin{figure}
\begin{center}
\includegraphics[angle=0,width=80mm]{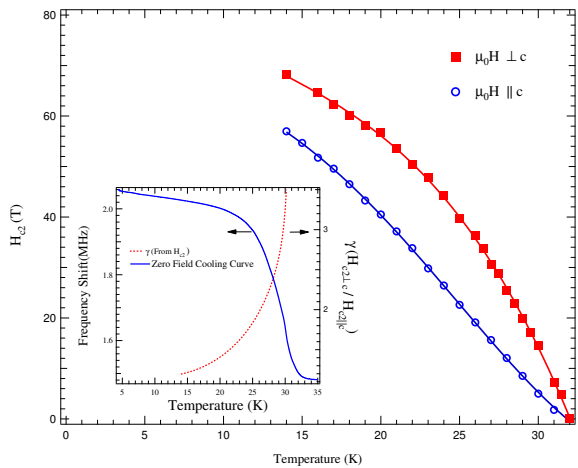}
\end{center}
\caption{(Color online) Anisotropic $H_{c2}(T)$ for (Ba$_{0.55}$K$_{0.45}$)Fe$_2$As$_2$ single crystals. Circles indicate the $H_{c2}^{ \parallel c}$ and squares $H_{c2}^{ \perp c}$. Inset:The rf zero field cooling curve for (Ba$_{0.55}$K$_{0.45}$)Fe$_2$As$_2$ single crystals on left axis and temperature-dependent anisotropy of $H_{c2}$: $\gamma = H_{c2}^{\perp c}/H_{c2}^{\| c}$ as determined from the fits to the $H_{c2}$ data on right axis} \label{F3}
\end{figure}

The shapes of the upper critical field curves for the parallel and perpendicular orientations clearly do not manifest the same temperature dependence . The anisotropy parameter $\gamma$ (inset to Fig. 3) varies accordingly. Estimates of the anisotropic coherence length at 14 K, based solely on the vortex flux density argument, yield $\xi_{\perp c}(0)\approx 34~\AA$ (in plane $\xi$) and $\xi_{\| c}(0) \approx 29~\AA$ (inter layer $\xi$) since $\xi_{\perp c}=\sqrt{\phi_{0}/2\pi H_{c2}^{\| c}}$ and $\xi_{\| c}=\phi_{0}/2\pi\xi_{\perp c}H_{c2}^{\perp c}$.\cite{hc2note} Decoupling of the superconducting layers would occur when $\xi_{\| c}(T)$ becomes less than the c-axis spacing, $c\approx 13~\AA$ for (Ba$_{0.55}$K$_{0.45}$)Fe$_{2}$As$_{2}$.\cite{ni}  To estimate the zero temperature coherence length, we note that $\xi(T)$ is approximated by G-L theory, i.e. $\xi(T) \propto \xi(0) / (1-t)^{1/2}$ where $t=T/T_c$. From $\xi_{\| c}(14 K)$  above, $\xi_{\| c}(0 K)$ is about $21.5 \AA$. Hence although the critical field anisotropy appears to decrease at lower temperatures, there is no simple argument based on estimates of the temperature dependent coherence lengths to support a dimensional crossover in the temperature range of this experiment.  Increased magnetic field intensities exceeding 75 tesla will be required to fully map out the lower temperature critical field phase diagram. 

The anisotropy of $H_{c2}$ in (Ba$_{0.55}$K$_{0.45}$)Fe$_{2}$As$_{2}$ is also reported in reference \cite{j} by measurement of the electrical resistivity of samples from a different origin.  The work presented here differs from \cite{j} in that a double hump appears in the cooling data of reference \cite{j}.  This artifact may be due to multiple crystalline phases or a doping inconsistency. Our crystals manifest a single and continuous superconducting transition (shown in inset to Fig.3).

To summarize, we have measured the anisotropic $H_{c2}(T)$ curves for $(Ba_{0.55}K_{0.45})Fe_2As_2$ down to 14 K and up to 60 T. We find (i) a non linear temperature dependence of the $H_{c2}$ anisotropy of (Ba$_{0.55}$K$_{0.45}$)Fe$_{2}$As$_{2}$, although there is at present no direct evidence for superconducting layer decoupling, and (ii) $H_{c2} (T=0)$ for this compound may easily approach fields of 75 tesla or above. Higher field measurements will be required to
determine the lower temperature parts of the full temperature dependence of
these remarkably large $H_{c2}(T)$ curves.  On the other hand, perhaps even
further physical insight will be drawn from careful studies of the angular
dependence of $H_{c2}(T)$ for fields below 50 T.  Rotations from $H \parallel c$ to $H \parallel
[100]$ and $H \parallel [110]$, as well as in plane rotations studies should reveal a
wealth of details about this modest, and temperature dependent upper
critical field anisotropy.\\

\begin{acknowledgments}
We would like to thank Prof. James Brooks for valuable scientific discussion. Work at the Ames Laboratory was supported by the US Department of Energy - Basic Energy Sciences under Contract No. DE-AC02-07CH11358.  MA is supported in part by   NSF-DMR 0602859 through the research grant of Professor James Brooks of Florida State University.  KC is supported in part by the NNSA/DoE research grant DOE-DE-FG52-06NA26193 of  Dr. Stan Tozer of Florida State University. Work at Los Alamos National Laboratory, National High Magnetic Field Laboratory-Pulsed Field Facility was supported by NSF Division of Materials Research through DMR-0654118, the U.S. Department of Energy and the State of Florida. Advances in contactless penetration depth technology is credited to Los Alamos National Laboratory LDRD-DR20070013. Los Alamos National Laboratory is operated by LANS LLC.
\end{acknowledgments}


\begin{thebibliography}{99}
\bibitem{hc2note} The estimate for $H_{c2}^{\perp c}(14)$ is 68 tesla and for $H{c2}^{\| c}(14)$ is 57 tesla.
\end{thebibliography}
\end{document}